\documentclass{new_tlp}
\usepackage{yfonts}
\usepackage{url}
\def\gringo{{\sc gringo}}

\def\gm{\textfrak{M}}
\def\uc{\tilde\forall}
\def\seq{\Rightarrow}
\def\Int{\emph{Int\/}}
\def\HTA{\emph{HTA\/}}

\def\num{\overline}

\def\ar{\leftarrow}
\def\rar{\rightarrow}
\def\lrar{\leftrightarrow}

\def\beq{\begin{equation}}
\def\eeq#1{\label{#1}\end{equation}}
\def\ba{\begin{array}}
\def\ea{\end{array}}
\def\ii#1{\hbox{\it #1\/}}
\def\no{\ii{not}}

\def\p2f{\hbox{p2f}}
\def\G{\Gamma}
\def\D{\Delta}
\def\r#1#2{\frac{\textstyle #1}{\textstyle #2}}

\title{Here and There with Arithmetic}

\author[Vladimir Lifschitz]{
  Vladimir Lifschitz\\
  University of Texas at Austin, USA
}

\begin{document}
\maketitle

\begin{abstract}
In the theory of answer set programming, two groups of rules are called
strongly equivalent if, informally speaking, they have the same meaning in
any context.  The relationship between strong equivalence and the
propositional logic of here-and-there allows us
to establish strong equivalence by deriving rules of each group from rules
of the other.  In the process, rules are rewritten
as propositional formulas.  We extend this
method of proving strong equivalence to an answer set programming
language that includes operations on integers.
The formula representing a rule in this language is
a first-order formula that may contain comparison symbols among its
predicate constants, and symbols for arithmetic operations among its
function constants.
The paper is under consideration for acceptance in TPLP.
\end{abstract}
  
\section{Introduction}

In the theory of answer set programming, two groups of rules are called
strongly equivalent if, informally speaking, they have the same meaning in
any context.  The relationship between strong equivalence and the
propositional logic of here-and-there~\cite{lif01} allows us
to establish strong equivalence by deriving rules of each group from rules
of the other.  In the process, rules are rewritten
as propositional formulas.  The head and the body of a rule become the
consequent and the antecedent of an implication; negation-as-failure is
replaced by classical negation; and choice expressions are tranformed into
excluded middle formulas. For instance, we rewrite the rule
\beq
\verb|{q} :- p|
\eeq{r1}
as the formula
\beq
p\rar q\lor\neg q,
\eeq{f1}
and the rule
\beq
\verb|q :- p, not not q|
\eeq{r2}
as the formula
\beq
p\land \neg\neg q \rar q.
\eeq{f2}
Formulas~(\ref{f1}) and~(\ref{f2}) can be derived from each other using the
postulates of intuitionistic logic and the Hosoi axiom schema
\beq
F\lor (F\rar G)\lor \neg G,
\eeq{hosoi}
which is part of the standard axiomatization of the logic  of here-and-there
\cite{hos66}.
We can conclude then that rules~(\ref{r1}) and~(\ref{r2}) are
strongly equivalent to each other.

Our goal is to extend this method of proving strong equivalence to
rules of mini-\gringo---the fragment of the input language of
\gringo\ \cite{gringomanual}
investigated in several recent publications
\cite{lif19,fan20,lif20,lif21}.  Since mini-\gringo\ rules may contain
variables, comparison symbols, and
symbols for arithmetic operations, the formula representing a rule will be
a first-order formula that may contain comparison symbols among its
predicate constants, and symbols for arithmetic operations among its
function constants.

The study of strong equivalence for fragments of the
input language of \gringo\ may be useful for the practice of answer set
programming, because it can help the programmer to see which modifications of
a rule or a group of rules preserve the set of stable models.
This is important, in particular,
because some features of the strong equivalence relation
between \gringo\ programs may seem counterintuitive.  For instance, the rule
$$
\verb|p(1..X) :- p(1..X), q(X)|
$$
appears to be trivial and strongly equivalent to the empty program, but it is
not.  Adding this rule to
the pair of facts \verb|p(2)|, \verb|q(2)| gives the program with a
stable model that includes \verb|p(1)|.  (The reason is that \gringo\ treats
\verb|p(1..X)| in the head of a rule as a conjunction, and in the body as a
disjunction.)  To give another example, the one-rule programs
\beq
\verb|p(X,Y) :- X = 1..2, Y = 1..2|
\eeq{rule2}
and
\beq
\verb|p(X,Y) :- X = Y, Y = 1..2|
\eeq{rule3}
have different stable models.

Stable models of a mini-\gringo\ program~$\Pi$
are defined as stable models of the set of propositional formulas obtained
from~$\Pi$ by the syntactic transformation~$\tau$ \cite[Section~3]{lif19}.
The definition of~$\tau$ involves naive grounding, so that the set
$\tau\Pi$ is infinite whenever~$\Pi$ contains variables.
Mini-\gringo\ programs~$\Pi_1$ and~$\Pi_2$ are \emph{strongly equivalent} to
each other if~$\tau\Pi_1$ is strongly equivalent to~$\tau\Pi_2$, that is to
say, if, for every set~$\Omega$ of propositional formulas,
$\tau\Pi_1\cup\Omega$ has the same stable models as $\tau\Pi_2\cup\Omega$.
The paper \cite{lif19} where this definition was proposed aimed at verifying
strong equivalence of mini-\gringo\ programs using theorem provers
for classical logic.  Here we study the same equivalence relation, but the
goal is different: we would like to develop methods for
proving strong equivalence using non-classical reasoning ``manually,''
without assistance of automated reasoning tools.

The next section provides background information on the syntax of
mini-\gringo, on the logic of here-and-there, and on transforming
mini-\gringo\ rules into sentences of a first-order language.
% To represent rules by formulas, we use here two transformations:~$\tau^*$
%\cite[Section~6]{lif19} and~$\nu$ \cite{lif21}.
In Section~\ref{sec:hta} we describe the deductive system \HTA\ (for
``here-and-there with arithmetic''), which  can be used to establish strong
equivalence between mini-\gringo\ programs.  A theorem
expressing this fact is stated in Section~\ref{sec:proving} and proved in
Section~\ref{sec:proof}.  Then we show a pair of strongly
equivalent mini-\gringo\ programs to which the method proposed in this
paper is not applicable (Section~\ref{sec:counter}) and outline
directions for future work (Section~\ref{sec:conclusion}).

\section{Background}

\subsection{Syntax of Mini-GRINGO} \label{ssec:syntax}

Programs below are written in ``abstract syntax'' that disregards some
details related to representing programs by strings of ASCII characters
\cite{geb15}.
We assume that three countably infinite sets of symbols are selected:
\emph{numerals}, \emph{symbolic constants}, and \emph{variables}.
We assume that a 1-1 correspondence between numerals
and integers is chosen; the numeral corresponding to an integer~$n$ will be
denoted by $\num n$.
\emph{Precomputed terms} are numerals, symbolic constants, and the symbols \emph{inf}, \emph{sup}.  Terms allowed in a mini-\gringo\ program are formed from
precomputed terms and
variables using the six operation names
\beq
+\quad-\quad\times\quad/\quad\backslash \quad..
\eeq{ops}

An \emph{atom} is a symbolic constant optionally followed by a tuple
of terms in parentheses.  A \emph{literal} is an atom
possibly preceded by one or two occurrences of \emph{not}. A \emph{comparison}
is an expression of the form
$t_1\prec t_2$, where $t_1$, $t_2$ are program terms and $\prec$ is one
of the six comparison symbols
$$
=\quad\neq\quad<\quad>\quad\leq\quad\geq
$$

A \emph{rule} is an expression of the form
\beq
\emph{Head}\ar\emph{Body},
\eeq{rule}
where
\begin{itemize}
\item
\strut\emph{Body} is a conjunction (possibly empty) of literals and comparisons,
and
\item
  \strut\emph{Head} is either an atom (then~(\ref{rule}) is a
  \emph{basic rule\/}), or an atom in braces
(then~(\ref{rule}) is a \emph{choice rule}\/), or empty (then~(\ref{rule})
is a \emph{constraint}).
\end{itemize}
A \emph{program} is a finite set of rules.

As mentioned in the introduction, the semantics of mini-\gringo\ is defined
by transforming rules into sets of propositional formulas
\cite[Section~3]{lif19}.

\subsection{Logic of Here and There} \label{ssec:ht}

\emph{(Propositional) formulas} are formed from propositional atoms and the
symbol $\bot$ (``false'') using the connectives $\land,\ \lor,\ \rar$;
$\neg F$ is shorthand for $F\rar\bot$, and $F\lrar G$ is shorthand for
$(F\rar G)\land(G\rar F)$.
The deductive systems discussed in this paper are based on the version of
natural deduction that operates with \emph{sequent\/}s---expressions
$\G\seq F$, in which $\G$ is a finite set of
formulas (``assumptions''), and~$F$ is a formula.  We write sets of
assumptions as lists.  For instance, $A_1,A_2\seq F$ is shorthand for
$\{A_1,A_2\}\seq F$, and $\G,A\seq F$ is shorthand for $\G\cup\{A\}\seq F$.
A sequent of the form $\seq F$ will be identified with the formula~$F$.

\begin{figure}
% \hrule\medskip
$$\begin{array}{ll}
\!(\land I)\;\r	{\G\seq F \quad \D\seq G}
		{\G,\D\seq F\land G}%\qquad\qquad
&\quad		
(\land E)\;\r	{\G\seq F \land G}
		{\G\seq F}
	\quad
	    \r  {\G\seq F \land G}
		{\G\seq G}
\\ \\
\!(\lor I)\;\r	{\G\seq F}
		{\G\seq F \lor G}\quad
	    \r	{\G\seq G}
		{\G\seq F \lor G}%\quad
&\quad				
(\lor E)\;\r{\G\seq F \lor G \quad \D_1,F \seq H \quad \D_2,G\seq H}
	    {\G,\D_1,\D_2\seq H}
\\ \\
\!(\rar\!\! I)\;\r	{\G,F\seq G}
			{\G\seq F\rar G}
&\quad				
(\rar\!\! E)\;\r	{\G\seq F\quad \D\seq F \rar G}
			{\G,\D\seq G}
\end{array}
$$
% \medskip\hrule\medskip
\caption{Introduction and elimination rules for propositional connectives.}
\label{fig1}
\end{figure}
The postulates of propositional intuitionistic logic are the axiom schema
$F\seq F$, the introduction and elimination rules shown in Figure~\ref{fig1},
the contradiction rule
$$
(C)\;\r {\G\seq\bot}{\G\seq F}
$$
and the weakening rule
$$
(W)\;\r {\G\seq\Sigma}{\G,\D\seq\Sigma}.
$$

Classical propositional logic can be formalized by adding the law of the
excluded middle $F\lor\neg F$ to this deductive system as another
axiom schema.  The logic of here-and-there is the result of
adding a weaker axiom schema, Hosoi axiom~(\ref{hosoi}).  Thus the logic
of here-and-there is intermediate between intuitionistic and
classical.\footnote{A semantics corresponding to this
  intermediate logic can be defined in terms
  of Kripke models with two worlds (``here'' and ``there'') or in terms
of three-valued truth tables.
These truth tables were originally introduced by Heyting~\citeyear{hey30}
as a technical device for the purpose of demonstrating that intuitionistic
logic is weaker than classical.  Heyting remarks that the truth values in
these tables ``can be interpreted as follows: 0 denotes a correct
proposition, 1 denotes a false proposition, and 2 denotes a proposition that
cannot be false but whose correctness is not proved.''}

The implication from~(\ref{f1}) to~(\ref{f2}) can be proved intuitionistically
(Figure~\ref{fig:ex}).
\begin{figure}
% \hrule\medskip
$$\ba{lll}
1. & p\rar q\lor\neg q\;\seq\; p\rar q\lor\neg q
   & \hbox{ --- axiom}\\
2. & p\land \neg\neg q\;\seq\; p\land \neg\neg q
   & \hbox{ --- axiom}\\
3. & p\land \neg\neg q\;\seq\; p
   & \hbox{ --- from 2 by }(\land E)\\
4. & p\land \neg\neg q\;\seq\; \neg\neg q
   & \hbox{ --- from 2 by }(\land E)\\
5. & p\rar q\lor\neg q,\, p\land \neg\neg q \;\seq\; q\lor\neg q
   & \hbox{ --- from 3 and 1 by }(\rar E)\\
6. & q\;\seq\; q
   & \hbox{ --- axiom}\\
7. & \neg q\;\seq\; \neg q
   & \hbox{ --- axiom}\\
8. & p\land \neg\neg q,\, \neg q\;\seq\; \bot
   & \hbox{ --- from 7 and 4 by }(\rar E)\\
9. & p\land \neg\neg q,\, \neg q\;\seq\; q
   & \hbox{ --- from 9 by }(C)\\
10. & p\rar q\lor\neg q,\, p\land \neg\neg q \;\seq\; q
    & \hbox{ --- from 5, 6 and 9 by }(\lor E)\\
11. & p\rar q\lor\neg q, \;\seq\;  p\land \neg\neg q\rar q
    & \hbox{ --- from 10 by }(\rar I)\\
12. & \seq\; (p\rar q\lor\neg q)\rar(p\land \neg\neg q\rar q)
    & \hbox{ --- from 11 by }(\rar I)
\ea$$
% \medskip\hrule\medskip
\caption{Intutionistic proof of the implication from (\ref{f1})
  to~(\ref{f2}).}
\label{fig:ex}
\end{figure}
Informally, this proof can be summarized as follows.  Assume
$p\rar q\lor\neg q$ and $p\land \neg\neg q$.  Then $p$ and consequently
$q\lor\neg q$.  The second disjunctive term
contradicts $\neg\neg q$; consequently~$q$.  We
have derived $p\land \neg\neg q\rar q$ from $p\rar q\lor\neg q$.

The implication from (\ref{f2}) to~(\ref{f1}) can be proved using the
instance
$$q\lor(q\rar\neg q)\lor\neg\neg q$$
of axiom schema~(\ref{hosoi}).  Informally, we consider the cases
corresponding to the disjunctive terms of this disjunction.
\emph{Case~1:} $q$.  Then $q\lor\neg q$, which implies~(\ref{f1}).
\emph{Case~2:} $q\rar\neg q$.  Assuming~$q$ leads to a contradiction;
consequently $\neg q$.  Then $q\lor\neg q$, which implies~(\ref{f1}).
\emph{Case~3:} $\neg\neg q$.  Assume~$p$.  Then, by~(\ref{f2}), $q$,
and consequently $q\lor\neg q$.  Thus~(\ref{f1}) follows in this case as well.

\subsection{Two-Sorted Formulas and the System Int}\label{ssec:review}

The target language of transformations that convert mini-\gringo\ rules into
sentences is a first-order language with two sorts: the sort \emph{generic}
and its subsort \emph{integer}.  Variables of the first sort are meant to
range over arbitrary precomputed terms, and we will identify them with
variables used in mini-\gringo\ rules.  Variables of the second sort are meant
to range over numerals (or, equivalently, integers).  The need to use a
language with two sorts is explained by the fact that function symbols
in a first-order language are supposed to represent total functions, and
arithmetic operations are not defined on symbolic constants.  We will use
letters~$X$, $Y$, $Z$ to represent variables of the sort \emph{generic},~$M$
and~$N$ for variables of the sort \emph{integer}, and~$V$ when the sort is
irrelevant.

Formulas of the two-sorted system of
intuitionistic logic \Int\ \cite[Section~3]{lif21} are first-order
formulas of the signature~$\sigma$ that includes
\begin{itemize}
\item all precomputed terms as object constants; an object constant
  is assigned the sort \emph{integer} iff it is a numeral;
\item the symbols~$+$, $-$ and~$\times$ as binary function constants;
  their arguments and values have the sort \emph{integer};
\item pairs $p/n$, where $p$~is a symbolic constant and $n$ is a
  nonnegative integer, as $n$-ary predicate constants;
\item the comparison symbols
  $=,\,\neq,\,<,\,>,\,\leq,\,\geq$
  as binary predicate constants.
\end{itemize}
An atomic formula $(p/n)(t_1,\dots,t_n)$ can be abbreviated as
$p(t_1,\dots,t_n)$. An atomic formula $\prec\!\!(t_1,t_2)$, where~$\prec$ is a
comparison symbol, can be written as $t_1\prec t_2$.  Conjunctions of
equalities and inequalities can be abbreviated as usual in algebra (for
instance, $X<Y=Z$ is shorthand for $X<Y \land Y=Z$).

Note the difference between the set of terms over~$\sigma$ and the
set of terms allowed in mini-\gringo\ programs.
The former includes terms containing
integer variables, which are not allowed in mini-\gringo.
On the other hand, the latter includes
terms containing the symbols $/$, $\backslash$ and $..$ that are not
part of~$\sigma$.\footnote{The division
  and remainder operations are not allowed even in application to
  integer terms because division by 0 is undefined.
  The interval construction operation is not allowed because an
  interval does not correspond to a numeral (or to any precomputed term,
  for that matter); it represents a \emph{set} of numerals.}

Formulas over~$\sigma$ are formed from atomic formulas, the symbols
$\bot$, $\land$, $\lor$, $\rar$ and the quantifiers $\forall$, $\exists$
as usual in first-order logic.
Derivable objects of \Int\ are sequents, formed from
formulas over~$\sigma$ as described in Section~\ref{ssec:ht} for the
propositional case.  The system has two axiom schemas, $F\seq F$ and $t=t$,
where~$t$ is a term over $\sigma$.  Its inference rules are the rules
listed in
Section~\ref{ssec:ht} and the additional rules for quantifiers and equality
shown in Figure~\ref{fig2}.
\begin{figure}
% \hrule\medskip
$$\begin{array}{ll}
\!(\forall I)\;\r	{\G \seq F}
		{\G \seq \forall V F}
&\quad
(\forall E)\;\r	{\G \seq \forall V F}
		{\G \seq F^V_t}
\\ \\
\hbox{where $V$ is not free in~$\G$}
&\quad\hbox{where $t$ is substitutable for $V$ in $F$}
\\ \\
\!(\exists I)\;\r	{\G \seq F^V_t}
		{\G \seq \exists V F}
&\quad
\!(\exists E)\;\r	{\G \seq \exists V F \quad \D,F \seq G}
		{\G,\D \seq G}
\\ \\
\hbox{where $t$ is substitutable for $V$ in $F$}
&\quad\hbox{where $V$ is not free in~$\D,G$}
\end{array}
$$
$$\ba c
(Eq)\quad
		\r {\G \seq t_1=t_2 \quad \D \seq F^V_{t_1}}
		   {\G,\D\seq F^V_{t_2}}
\quad
		\r {\G \seq t_1=t_2 \quad \D \seq F^V_{t_2}}
                {\G,\D\seq F^V_{t_1}}
\\ \\
                \hbox{where $t_1$ and $t_2$ are substitutable for $V$ in $F$}
\ea$$
\medskip

\caption{Additional rules for quantifiers and equality.
By $F^V_t$ we denote
for the formula obtained from~$F$ by substituting the term~$t$ for
all free occurrences of the variable~$V$.  In rules~$(\forall E)$,
$(\exists I)$ and~$(Eq)$
the terms substituted for~$V$ are required to be of the sort \emph{integer}
if~$V$ is of the sort \emph{integer}.}
\label{fig2}
\end{figure}

\subsection{Transforming Rules into Formulas}

The translation~$\tau^*$ \cite[Section~6]{lif19} transforms mini-\gringo\
rules into sentences over the signature~$\sigma$ defined above.  In
some cases, it converts a short rule into a syntactically complex formula
that can be made much shorter by intuitionistically equivalent
transformations.  This is related to the generality of
the concept of a term in the input language of \gringo.  Its syntax allows
us to combine applications of the interval operation $..$ with addition and
multiplication in any order.  This feature is preserved in mini-\gringo;
we can write, for instance,
$((\num 1\,..\,\num 3)\times(\num 4\,..\,\num 6))+(\num 7\,..\,\num 9)$,
and even $(\num 1\,..\,\num 3)\,..\,\num 4$.  (The latter actually has
the same meaning as $\num 1\,..\,\num 4$.)  The need to translate rules that
contain combinations like these may cause significant overhead when~$\tau^*$
is applied to less exotic rules.

This overhead can be avoided if we do not try to cover arbitrary rules
\cite{lif21}.  A mini-\gringo\ term is \emph{a regular term of
the first kind} if
\begin{itemize}
\item it contains no operation symbols other than $+$, $-$, $\times$, and
\item symbolic constants and the symbols \emph{inf}, \emph{sup} do not
  occur in it in the scope of operation symbols.
\end{itemize}
\emph{A regular term of the second kind\/} is a term of
the form $t_1\,..\,t_2$, where $t_1$ and $t_2$ are regular terms
of the first kind that contain neither symbolic constants nor the
symbols \emph{inf}, \emph{sup}.  A rule~(\ref{rule}) is
\emph{regular} if it satisfies the following conditions:
\begin{itemize}
\item every term occurring in it is regular (of the first or second kind),
\item if a conjunctive term of \emph{Body} is a literal then it does not
contain terms of the second kind,
\item if a conjunctive term of \emph{Body} is a comparison that contains a
  term of the second kind then it has the form $t_1=t_2\,..\,t_3$,
  where~$t_1$ is a term of the
  first kind different from symbolic constants and from the symbols
  \emph{inf}, \emph{sup}.
\end{itemize}

The translation~$\nu$ transforms every regular rule~$R$ into a
sentence over~$\sigma$ such that the equivalence between $\nu R$ and
$\tau^*R$ is provable in \Int.
We review here the definition of~$\nu$ for the special case when the head of
the rule does not contain terms of the second kind \cite[Section~4]{lif21}.
Applying the translation~$\nu$ to such a rule
involves substituting variables of the sort \emph{integer}
for the variables
that occur in that rule at least once
in the scope of an operation symbol or in a comparison of the second kind.
Make the list $X_1,\dots,X_m$ of all such variables, and
choose~$m$ distinct \emph{integer} variables
$N_1,\dots,N_m$.  Substituting $N_1,\dots,N_m$ for $X_1,\dots,X_m$ in a
term~$t$ of the first kind that occurs in~$R$ eliminates \emph{generic}
variables in the scope of operation symbols, so that the result of
this substitution is a term over~$\sigma$.  This term is denoted by $\p2f(t)$,
and similarly for the result of this substitution applied to a tuple of terms.

The result of applying~$\nu$ to a rule~(\ref{rule}) is defined as
the universal closure of the formula
$$\emph{Body\/}'\rar\emph{Head\/}',$$
where $\emph{Body\/}'$ is obtained from \emph{Body} by replacing
\begin{itemize}
\item every occurence of an atom $p(\bf t)$ by $p(\p2f({\bf t}))$,
\item every occurence of $\no$ by $\neg$,
\item every comparison $t_1\prec t_2$ of the first kind by
  $\p2f(t_1)\prec\p2f(t_2)$,
\item every comparison $t_1\prec t_2\,..\,t_3$ of the second kind by
  $\p2f(t_2)\leq\p2f(t_1)\leq\p2f(t_3)$;
\end{itemize}
and $\emph{Head\/}'$ is
\begin{itemize}
\item $p(\p2f({\bf t}))$ if \emph{Head} is $p(\bf t)$,
\item $p(\p2f({\bf t}))\lor\neg p(\p2f({\bf t}))$ if \emph{Head} is
  $\{p(\bf t)\}$,
\item $\bot$ if \emph{Head} is empty.
\end{itemize}

For instance,
% the result of applying~$\nu$ to the rule
% $$q(X,Y,Z) \ar p(X) \land Y = \num 1\,..\,\num 3 \land Z = X+\num 1$$
% is the formula
% $$\forall N_1N_2Z(p(N_1)\land \num 1\leq N_2\leq \num 3 \land
% Z=N_1+\num 1 \rar q(N_1,N_2,Z)).$$
% Rule~(\ref{rule1}) is not regular.
rules~(\ref{rule2}) and~(\ref{rule3})
are regular, and the transformation~$\nu$ turns them into the formulas
$$\ba c
\forall MN(\num 1\leq M\leq \num 2 \land \num 1\leq N\leq \num 2
\rar p(M,N)),\\
\forall XN(X=N \land \num 1\leq N\leq \num 2 \rar p(X,N)).
\ea$$

The result of applying~$\tau^*$ to a program~$\Pi$ is the set (or conjunction)
of the sentences $\tau^*R$ for all rules~$R$ of~$\Pi$, and similarly for~$\nu$.

\section{Deductive System HTA} \label{sec:hta}

The system \HTA\ is obtained from the system \Int\ reviewed in
Section~\ref{ssec:review} by adding the following postulates.
\begin{enumerate}
\item[A.]
  Hosoi axiom schema~(\ref{hosoi}).
\item[B.]
  The decidability axioms
  \beq
  X\prec Y \lor\neg\,(X\prec Y)
  \eeq{emcomp}
  for all comparison symbols $\prec$,
and the axioms expressing that comparison symbols describe a total order
  with the smallest element \emph{inf} and the largest element \emph{sup}.
  Some classically equivalent formulas expressing properties of order
  relations are not equivalent intuitionistically; for instance
  $$X\leq Y \land X\neq Y \rar X < Y$$
  is weaker, intuitionistically, than
  $$X\leq Y \rar X < Y \lor X=Y.$$
  But in the presence of decidability axioms~(\ref{emcomp}) such formulas are
  interchangeable.  For this reason, there is no need to go into details
  about the choice of axioms in this group.
\item[C.]
  All true formulas of the forms $c_1\prec c_2$ and $\neg\,(c_1\prec c_2)$,
  where~$c_1$, $c_2$ are precomputed terms and $\prec$ is a
  comparison symbol.  If, for instance, strings of lowercase letters are
  symbolic constants and they are ordered lexicographically then
  $\emph{ab}<\emph{ac}$ is one of the axioms in this group.
\item[D.]
  All arithmetical sentences that are true in the domain of integers.
  (We call a formula arithmetical if all its atomic subformulas are
  comparisons between \emph{integer} terms.)  Examples:
  $$\num 2 \times \num 2 = \num 4,\
  \forall N(N\times N\geq \num 0).$$
\end{enumerate}

\section{Proving Strong Equivalence}\label{sec:proving}

\noindent{\bf Theorem.} \emph{
  For any mini-\gringo\ programs~$\Pi_1$ and~$\Pi_2$, if the equivalence
  between $\tau^*\Pi_1$ and $\tau^*\Pi_2$ can be proved in HTA then~$\Pi_1$
  is strongly equivalent to~$\Pi_2$.
}

\medskip
Since the equivalence between $\tau^*\Pi_i$ and $\nu\Pi_i$ is provable in
\Int\ when all rules are regular, the
translation~$\tau^*$ can be replaced by~$\nu$ in this special case:

\medskip\noindent{\bf  Corollary.} \emph{
  For any mini-\gringo\ programs~$\Pi_1$ and~$\Pi_2$ that consist of
  regular rules, if the equivalence between $\nu \Pi_1$ and $\nu \Pi_2$
  can be proved in HTA then~$\Pi_1$ is strongly equivalent to~$\Pi_2$.
}

\medskip
The examples below show how the corollary can be used to prove strong
equivalence of mini-\gringo\ programs.

\medskip\noindent{\bf Example 1.}
The rule
\beq
\{q(X)\} \ar p(X+\num 1)
\eeq{ex0l}
is strongly equivalent to
\beq
q(X) \ar p(X+\num 1) \land \no\ \no\ q(X),
\eeq{ex0r}
because the result
$$\forall N(p(N+\num 1)\rar q(N)\lor\neg q(N))$$
of applying~$\nu$ to~(\ref{ex0l}) is
equivalent in \HTA\ to the result
$$\forall N(p(N+\num 1) \land \neg \neg q(N) \rar q(N))$$
of applying~$\nu$ to~(\ref{ex0r}).  The derivation of this equivalence
is parallel to the derivation of the equivalence between~(\ref{f1})
and~(\ref{f2}) presented in Section~\ref{ssec:ht}, with the addition of
$\forall$-elimination and $\forall$-introduction steps.

\medskip\noindent{\bf Example 2.}
The program
\beq\ba {rl}
q(X)\!\!\!\! &\ar p(X),\\
q(X+\num 1)\!\!\!\! &\ar p(X+\num 1)
\ea\eeq{ex1}
is strongly equivalent to its first rule.  Indeed,~$\nu$ transforms
rules~(\ref{ex1}) into the formulas
\beq\ba l
\forall X(p(X)\rar q(X)),\\
\forall N (p(N+\num 1) \rar q(N+\num 1)).
\ea\eeq{two}
The second formula can be derived from the
first in \Int\ (by $\forall$-elimination followed by $\forall$-introduction).

\medskip
An attempt to prove the conjecture that
program~(\ref{ex1}) is strongly equivalent to its second rule fails.
(From the second of formulas~(\ref{two}) we can derive
$\forall N (p(N) \rar q(N))$ by substituting $N-\num 1$ for~$N$, but
the $\forall$-elimination rule of \Int\ does not sanction substituting~$X$
for~$N$.)  That conjecture is actually incorrect: adding
rules~(\ref{ex1}) to the fact $p(a)$ gives a program with the stable
model $\{p(a),q(a)\}$; adding the second rule alone does not allow us to
conclude $q(a)$.

\medskip\noindent{\bf Example 3.}
The program
\beq\ba {rl}
\{q(X,Y)\} \!\!\!\! &\ar p(X,Y)\land X < Y,\\
\{q(X,X)\} \!\!\!\! &\ar p(X,X)
\ea\eeq{ex1l}
is strongly equivalent to the rule
\beq
\{q(X,Y)\} \ar p(X,Y)\land X\leq Y.
\eeq{ex1r}
Indeed,~$\nu$ transforms the first of rules~(\ref{ex1l}) into
\beq
\forall XY(p(X,Y)\land X < Y \rar q(X,Y)\lor\neg q(X,Y)),
\eeq{ex1lf}
and the second into
$$\forall X (p(X,X)\rar q(X,X) \lor \neg q(X,X)).$$
The last formula is intuitionistically equivalent to
\beq
\forall XY(p(X,Y)\land X = Y \rar q(X,Y)\lor\neg q(X,Y)).
\eeq{ex1lfa}
On the other hand, $\nu$ transforms rule~(\ref{ex1r}) into
\beq
\forall XY(p(X,Y)\land X \leq Y \rar q(X,Y)\lor\neg q(X,Y)).
\eeq{ex1rf}
The equivalence of~(\ref{ex1rf}) to the conjunction of~(\ref{ex1lf})
and~(\ref{ex1lfa}) is an intuitionistic consequence of the formula
$$\forall XY(X\leq Y \lrar X<Y\lor X=Y),$$
which follows from the axioms in Group~B.

\medskip\noindent{\bf Example 4.}
The rule
\beq
q(X+Y) \ar p(X) \land p(Y) \land X\leq Y
\eeq{ex2l}
is strongly equivalent to
\beq
q(X+Y) \ar p(X) \land p(Y).
\eeq{ex2r}
Indeed,~$\nu$ transforms rule~(\ref{ex2l}) into
\beq
\forall MN(p(M)\land p(N)\land M\leq N \rar q(M+N))
\eeq{ex2lf}
and rule~(\ref{ex2r}) into
$$%\beq
\forall MN(p(M)\land p(N) \rar q(M+N)).
$$%\eeq{ex2rf}
The last formula is equivalent in \HTA\ to
$$\forall MN(p(M)\land p(N) \land(M\leq N \lor N\leq M)\rar q(M+N)),$$
(using the axiom $\forall MN(M\leq N \lor N\leq M)$ from Group~D).
Consequently it is equivalent to the conjunction of~(\ref{ex2lf}) and
$$\forall MN(p(M)\land p(N)\land N\leq M \rar q(M+N)).$$
It remains to observe that the last formula is equivalent to~(\ref{ex2lf})
by the commutativity of addition, which is an axiom in Group~D.
\medskip

The conditions for strong equivalence given by the theorem and
the corollary are sufficient, but not necessary.  A counterexample proving
this fact is given in Section~\ref{sec:counter}.

\section{Proof of the Theorem} \label{sec:proof}

Prior to presenting the proof of the theorem from Section~\ref{sec:proving},
we review the infinitary logic of here-and-there, which plays a
large part in the proof.

\subsection{Review: Infinitary Logic of Here-and-There}

The syntax of infinitary propositional formulas \cite{tru12} can be
described as follows.  For every nonnegative integer~$r$,
\emph{(infinitary) formulas of rank~$r$} over a given set
of atomic propositions are defined recursively, as follows:
\begin{itemize}
\item every atomic proposition is a formula of rank~0,
\item if $\mathcal{H}$ is a set of formulas, and~$r$ is the smallest
nonnegative 
integer that is greater than the ranks of all elements of $\mathcal{H}$,
then $\mathcal{H}^\land$ and $\mathcal{H}^\lor$ are formulas of rank~$r$,
\item if $F$ and $G$ are formulas, and~$r$ is the smallest nonnegative
integer that is greater than the ranks of~$F$ and~$G$, then $F\rar G$ is a
formula of rank~$r$.
\end{itemize}
We write $\{F,G\}^\land$ as $F\land G$, and $\{F,G\}^\lor$ as $F\lor G$.
The symbol $\bot$ is understood in this language as shorthand
for~$\emptyset^{\land}$, and~$\top$ is shorthand
for~$\emptyset^{\lor}$.  These conventions allow us to treat
propositional combinations of atomic propositions as a special case of
infinitary formulas.

For any family $\{F_\alpha\}_{\alpha\in A}$ of formulas whose ranks are
bounded, we denote the formula $\{F_\alpha:\alpha\in A\}^\land$ by
$\bigwedge_{\alpha\in A}F_\alpha$, and similary for disjunctions.

An \emph{interpretation} is a set of atomic propositions.
The satisfaction relation between an interpretation and a formula is
defined recursively, as follows:
\begin{itemize}
\item For every atomic proposition $p$, $I\models p$ if $p\in I$.
\item $I\models\mathcal{H}^\land$ if for every formula $F$ in~$\mathcal{H}$,
$I\models F$.
\item $I\models\mathcal{H}^\lor$ if there is a formula $F$ in~$\mathcal{H}$
such that $I\models F$.
\item $I\models F\rar G$ if $I\not\models F$ or $I\models G$.
\end{itemize}
% A {\sl model} of a set $\mathcal{H}$ of infinitary formulas is an
% interpretation that satisfies all formulas in~$\mathcal{H}$.
A formula is \emph{tautological} if it is satisfied by all interpretations.

% The definition of the strong equivalence relation
% between propositional formulas is extended to the infinitary case in a
% straightforward way \cite{har17}.

An \emph{HT-interpretation} is an ordered pair
$\langle I,J\rangle$ of interpretations such that
$I\subseteq J$.
The satisfaction relation between an HT-interpretation and an
infinitary formula~\cite{har17} is defined recursively, as follows:
\begin{itemize}
\item For every atomic proposition $p$, $\langle I,J\rangle\models p$ if
$p\in I$.
\item $\langle I,J\rangle\models\mathcal{H}^\land$ if for every formula
$F$ in~$\mathcal{H}$, $\langle I,J\rangle\models F$.
\item $\langle I,J\rangle\models\mathcal{H}^\lor$ if there is a formula
$F$ in~$\mathcal{H}$ such that $\langle I,J\rangle\models F$.
\item $\langle I,J\rangle\models F\rar G$ if
\begin{enumerate}
\item[(i)]
 $\langle I,J\rangle\not\models F$ or $\langle I,J\rangle\models G$, and
\item[(ii)]
 $J\models F\rar G$.
\end{enumerate}
\end{itemize}
% An {\sl HT-model} of a set $\mathcal{H}$ of infinitary formulas is an
% HT-interpretation that satisfies all formulas in~$\mathcal{H}$.

\subsection{Plan of the Proof}

A \emph{precomputed atom} is a
symbolic constant optionally followed by a tuple of precomputed terms
in parentheses.  The proof of the theorem uses the transformation
$F\mapsto F^{\rm prop}$ \cite[Section~5]{lif19} that converts every
sentence over the signature~$\sigma$ (Section~\ref{ssec:review}) into an
infinitary formula over the set of precomputed atoms.  The transformation
is defined as follows:
\begin{itemize}
\item
$p({\bf t})^{\rm prop}$ is obtained from~p({\bf t}) by replacing
each member of~${\bf t}$ by its value;
\item
$(t_1\prec t_2)^{\rm prop}$ is $\top$ if the
values of $t_1$ and $t_2$ are in the relation~$\prec$, and $\bot$
otherwise;
\item 
$\bot^{\rm prop}$ is $\bot$;
\item
$(F\odot G)^{\rm prop}$ is $F^{\rm prop}\odot G^{\rm prop}$ for every binary
connective~$\odot$;
\item
  $(\forall V F)^{\rm prop}$ is the conjunction of the formulas
$(F^V_r)^{\rm prop}$ over all precomputed terms $r$ if $V$ is \emph{generic},
and over all numerals~$r$ if $V$ is \emph{integer};
\item
$(\exists V F)^{\rm prop}$ is the disjunction of the formulas
$(F^V_r)^{\rm prop}$ over all precomputed terms $r$ if $V$ is \emph{generic},
and over all numerals~$r$ if $V$ is \emph{integer}.
\end{itemize}

\noindent{\bf Lemma 1} \cite[Proposition 4]{lif19}.
\emph{Mini-\gringo\ programs $\Pi_1$, $\Pi_2$ are strongly equivalent to
  each other iff the infinitary formula $(\tau^*\Pi_1)^{\rm prop}$ is
  strongly equivalent to $(\tau^*\Pi_2)^{\rm prop}$.
}

\medskip\noindent{\bf Lemma~2} \cite[Theorem~3]{har17}.
\emph{
Two infinitary formulas are strongly equivalent to each other
iff they are satisfied by the same HT-interpretations.
}
\medskip

  We extend the definition of
$F^{\rm prop}$ to sequents as follows:
$$(\G\seq F)^{\rm prop}=(\uc(\G^\land\rar F))^{\rm prop}.$$
(Here $\G^\land$ is the conjunction of all formulas in~$\G$;
$\uc$ is the universal closure operation).

\medskip\noindent{\bf Lemma 3.}
\emph{
  If a sequent~$S$ can be derived from sequents $S_1,\dots,S_k$ by one
  application of
  an inference rule of Int then every HT-interpretation satisfying
  $(S_1)^{\rm prop},\dots,(S_k)^{\rm prop}$ satisfies $S^{\rm prop}$.
}

\medskip\noindent{\bf Lemma 4.}
\emph{
  For every axiom~$S$ of HTA, $S^{\rm prop}$ is satisfied by all
  HT-interpretations.
}
\medskip

To derive the theorem from the lemmas, we reason as follows.  From
Lemmas~3 and~4 we can conclude that if a sequent~$S$ can be derived from
sequents~$S_1,\dots,S_k$ in \HTA\ then every HT-interpretation satisfying
$(S_1)^{\rm prop},\dots,(S_k)^{\rm prop}$ satisfies $S^{\rm prop}$.
In particular, the assumption that
$\tau^*\Pi_1$ and $\tau^*\Pi_2$ can be derived from each other
in \HTA\ implies that $(\tau^*\Pi_1)^{\rm prop}$ and
$(\tau^*\Pi_2)^{\rm prop}$ are satisfied by the same HT-interpretations.
By Lemma~2, it follows then that
$(\tau^*\Pi_1)^{\rm prop}$ is strongly equivalent to
$(\tau^*\Pi_2)^{\rm prop}$.  Then, by Lemma~1, $\Pi_1$ is strongly
equivalent to~$\Pi_2$.

To complete the proof of the theorem, we need to prove Lemmas~3 and~4.

\subsection{Proof of Lemma~3 (sketch)}

We show here how to prove the assertion of the lemma for two of the
inference rules of \Int.  One is implication elimination:
$$
(\rar\!\! E)\;\r{\G\seq F\quad \D\seq F \rar G}{\G,\D\seq G}.
$$
We need to show that every HT-interpretation satisfying the formulas
\beq
(\uc(\G^\land) \rar F)^{\rm prop}
\hbox{ and }
(\uc(\D^\land) \rar (F\rar G))^{\rm prop}
\eeq{ie17}
satisfies
\beq
(\uc((\G\cup\D)^\land \rar G))^{\rm prop}.
\eeq{ie2}
Let~$\bf V$ be the list $V_1,V_2,\dots$ of variables that
are free in $\G$, $\D$, $F$, or~$G$.  Formulas~(\ref{ie17}) can be written as
$$
\bigwedge_{\bf r}\left(\left((\G^\land)^{\bf V}_{\,\bf r}\right)^{\rm prop}
\rar \left(F^{\bf V}_{\,\bf r}\right)^{\rm prop}\right)
$$
and
$$\bigwedge_{\bf r}\left(\left((\D^\land)^{\bf V}_{\,\bf r}\right)^{\rm prop}
\rar \left(\left(F^{\bf V}_{\,\bf r}\right)^{\rm prop}\rar \left(G^{\bf V}_{\,\bf r}\right)^{\rm prop}\right)\right),
$$
where $\bf r$ ranges over the tuples $r_1,r_2,\dots$ of precomputed terms
such that $r_i$ is a numeral whenever the sort of~$V_i$ is \emph{integer}.
Similarly,~(\ref{ie2}) can be written as
$$
\bigwedge_{\bf r}\left(\left((\G^\land)^{\bf V}_{\,\bf r}\right)^{\rm prop}
  \land \left((\D^\land)^{\bf V}_{\,\bf r}\right)^{\rm prop}
\rar \left(G^{\bf V}_{\,\bf r}\right)^{\rm prop}\right)
$$
It remains to observe that if an HT-interpretation satisfies
$$
\left((\G^\land)^{\bf V}_{\,\bf r}\right)^{\rm prop} \rar
      \left(F^{\bf V}_{\,\bf r}\right)^{\rm prop}
$$
and
$$
\left((\D^\land)^{\bf V}_{\,\bf r}\right)^{\rm prop} \rar \left(\left(F^{\bf V}_{\,\bf r}\right)^{\rm prop}\rar \left(G^{\bf V}_{\,\bf r}\right)^{\rm prop}\right)
$$
then it satisfies
$$
\left((\G^\land)^{\bf V}_{\,\bf r}\right)^{\rm prop}\land
\left((\D^\land)^{\bf V}_{\,\bf r}\right)^{\rm prop}
\rar \left(G^{\bf V}_{\,\bf r}\right)^{\bf prop}.$$

Second, consider the existental quantifier introduction rule for the sort
\emph{integer:}
$$
(\exists E)\;\r{\G \seq F^N_t}{\G \seq \exists N\,F},
$$

\smallskip\noindent
where $t$ is an \emph{integer} term substitutable for $N$ in $F$.
The proof uses the following fact: replacing an occurrence of a ground term
in a closed formula~$F$ by its value does not
change $F^{\rm prop}$.  This is easy to check by induction.

We need to show that every HT-interpretation satisfying the formula
\beq
(\uc(\G^\land \rar F^N_{\,t}))^{\rm prop}
\eeq{ex17}
satisfies
\beq
(\uc(\G^\land \rar \exists N F))^{\rm prop}.
\eeq{ex2}
Let~$\bf V$ be the list $V_1,V_2,\dots$ of variables that
are different from~$N$ and are free in $\G$, $F$, or~$t$.  Formula~(\ref{ex17})
can be written as
$$
\bigwedge_{{\bf r},n}\left(
  \left((\G^\land)^{{\bf V},N}_{\,{\bf r},\;\num n}\right)^{\rm prop}
  \rar
  \left((F^N_{\,t})^{{\bf V},N}_{{\,\bf r},\;\num n}\right)
     ^{\rm prop}\right),
   $$
where $\bf r$ ranges over the tuples $r_1,r_2,\dots$ of precomputed terms
such that $r_i$ is a numeral whenever the sort of~$V_i$ is \emph{integer},
and $n$ ranges over integers.  Similarly,~(\ref{ex2}) can be written as
$$
\bigwedge_{{\bf r},n}\left(
\left((\G^\land)^{{\bf V},N}_{\,{\bf r},\;\num n}\right)^{\rm prop}
\rar
\bigvee_m\left(F^{{\bf V},N}_{\;{\bf r},\;\num m}\right)^{\rm prop}\right)
$$
($m$ ranges over integers).  It is sufficient to show that for all $\bf r$,
$n$, if an HT-interpretation satisfies
\beq
\left(F^N_{\,t})^{{\bf V},N}_{{\,\bf r},\;\num n}\right)
^{\rm prop}
\eeq{a}
then it satisfies
\beq
\bigvee_m\left(F^{{\bf V},N}_{\;{\bf r},\;\num m}\right)^{\rm prop}.
\eeq{b}
Since $t$ is substitutable for $N$ in~$F$,
formula~(\ref{a}) can be also written as
$$\left(F^{{\bf V},N}_{\;{\bf r},\,\;t^{{\bf V},N}_{{\,\bf r},\;\num n}}
\right)^{\rm prop}.$$
This is the disjunctive term of~(\ref{b}) for which~$\num m$ is the
value of $t^{{\bf V},N}_{{\,\bf r},\;\num n}$.

\subsection{Proof of Lemma~4}

For the axiom schemas $F\seq F$ and $t=t$ of \Int, the assertion of the
lemma is
obvious.  For axioms of Group~A, it is verified in the same way as the
validity of Hosoi axioms in the logic of here-and-there.

The images of the axioms of Groups B--D under the transformation
\hbox{$F\mapsto F^{\rm prop}$} belong to the class of negative infinitary
formulas, which is defined recursively:
\begin{itemize}
\item $\mathcal{H}^\land$ and $\mathcal{H}^\lor$ are negative if every formula
in~$\mathcal{H}$ is negative;
\item $F\rar G$ is negative if~$G$ is negative.
\end{itemize}
Every negative tautological formula is satisfied by all
HT-interpretations \cite[Theorem~1]{har17}.  Consequently
we only need to check that the images of all axioms of Groups B--D are
tautological.  For Groups~B and~C, this property is obvious.  For
Group~D, it is easy to check by induction that for any arithmetical
sentence~$F$,
$F^{\rm prop}$ is tautological if~$F$ is true in the domain of integers, and
$\neg F^{\rm prop}$ is tautological otherwise.
  
\section{HTA Is Incomplete for Strong Equivalence}\label{sec:counter}

The sufficient conditions for strong equivalence of mini-\gringo\
programs given by the theorem from Section~\ref{sec:proving} and its
corollary are not necessary.  Indeed, consider the program~$\Pi_1$
consisting of the rules
$$
\ba l
p(\num 0),\\
p(X+\num 1) \ar p(X)
\ea
$$
and the program~$\Pi_2$, obtained from~$\Pi_1$ by adding the rule
\beq
p(X) \ar X+\num 1 > \num 0.
\eeq{ce2}
These programs are strongly equivalent to each other.  Indeed, the
transformation~$\tau$  \cite[Section~3]{lif19} converts~$\Pi_1$
into a set of formulas that contains
$p(\num 0)$ and the implications
$$p(\num n) \rar p(\num{n+1})$$
for all integers~$n$.  Rule~(\ref{ce2}) is transformed by~$\tau$ into the set
consisting of (i)~the formulas
\beq
\top\rar p(\num{n})
\eeq{ce2a}
for all nonnegative integers~$n$, and~(ii) some implications with the
antecedent~$\bot$.\footnote{Implications with the
  antecedent~$\bot$ correspond to the ground instances of~(\ref{ce2}) in
  which the precomputed term substituted for~$X$ is not a numeral or is
  less than~$\num 0$.}
 It is clear that~$\tau\Pi_2$
can be derived from~$\tau\Pi_1$ in propositional intuitionistic logic.

We will show now that the formula
\beq
\forall N(N+\num 1 > \num 0 \rar p(N)),
\eeq{ce2b}
which is obtained from rule~(\ref{ce2}) by applying the transformation~$\nu$,
cannot be derived in \HTA\ from the formulas
\beq\ba l
p(\num 0),\\
\forall N(p(N)\rar p(N+\num 1)),
\ea\eeq{ce1}
which are obtained by applying~$\nu$ to the rules of~$\Pi_1$.
In fact, formula~(\ref{ce2b}) cannot be derived from $\nu\Pi_1$ even in the
classical theory obtained from \HTA\ by replacing the Hosoi axiom schema
(Group~A) by the law of the excluded middle.  Call this classical
first-order theory~$T$.  Let~$\gm$
be a model of~$T$ in which the integer universe is nonstandard,
that is, has an element that does not correspond to a numeral.
Since the sentence
$$\forall N (N\geq\num 0 \lor -N\geq\num 0)$$
is an axiom of~$T$, the integer universe of $\gm$ has an element
that does not correspond to a numeral and is nonnegative; call it~$\alpha$.
Let~$\gm'$ be obtained from~$\gm$ by reinterpreting~$p$ so that the extent
of~$p$ in~$\gm'$ is the set of objects represented by nonnegative numerals.
Since~$p$ does not occur in any of the axioms B--D, $\gm'$ is a model of~$T$
as well.  It is clear that~$\gm'$ satisfies~(\ref{ce1}) but does not
satisfy~(\ref{ce2b}): $\alpha$ is a counterexample.

\section{Future Work}\label{sec:conclusion}

The incompleteness of \HTA\ as a tool for establishing strong equivalence
of mini-\gringo\ programs suggests one direction for future work: making
\HTA\ stronger.  Adding an induction axiom schema that is not restricted to
arithmetical formulas may allow us to handle the example of
programs~$\Pi_1$,~$\Pi_2$ from Section~\ref{sec:counter}.  Another possible
addition to \HTA\ is the axiom schema
$$\exists V(F \rar \forall VF),$$
familiar from research on strong equivalence of logic programs with
variables  \cite[Section~3]{lif07a}.  Will these or similar additions
make \HTA\ complete for strong equivalence in mini-\gringo?

On the other hand, it is not clear how useful such additional axioms are
going to be from the programmer's perspective.  The stable model
of~$\Pi_1$,~$\Pi_2$ is infinite; these programs cannot be processed by
grounders, such as \gringo.  Do there exist similar examples with finite
stable models?

It would be perhaps more useful to extend the syntax of mini-\gringo\ by
aggregates, such as counting, which play an important role in the practice
of answer set programming, and then extend the translations~$\tau^*$
and~$\nu$ and the system \HTA\ accordingly.

\section*{Acknowledgements}

Thanks to David Pearce for insightful comments related to the topic of this
paper, and to the anonymous referees for good advice on improving
presentation.

\bibliographystyle{acmtrans}
\bibliography{bib}
\end{document}